The New Horizons Spacecraft

Glen H. Fountain, David Y. Kusnierkiewicz, Christopher B. Hersman, Timothy S. Herder, Thomas B Coughlin,  William T. Gibson, Deborah A. Clancy, Christopher C. DeBoy, T. Adrian Hill, James D. Kinnison, Douglas S. Mehoke, Geffrey K. Ottman, Gabe D. Rogers, S. Alan Stern, James M. Stratton, Steven R. Vernon, Stephen P. Williams

Abstract

The New Horizons spacecraft was launched on January 19, 2006. The spacecraft was designed to provide a platform for the seven instruments designated by the science team to collect and return data from Pluto in 2015 that would meet the requirements established by the National Aeronautics and Space Administration (NASA) Announcement of Opportunity AO-OSS-01. The design drew on heritage from previous missions developed at The Johns Hopkins University Applied Physics Laboratory (APL) and other NASA missions such as Ulysses.  The trajectory design imposed constraints on mass and structural strength to meet the high launch acceleration consistent with meeting the AO requirement of returning data prior to the year 2020.  The spacecraft subsystems were designed to meet tight resource allocations (mass and power) yet provide the necessary control and data handling finesse to support data collection and return when the one way light time during the Pluto fly-by is 4.5 hours.  Missions to the outer regions of the solar system (where the solar irradiance is 1/1000 of the level near the Earth) require a Radioisotope Thermoelectric Generator (RTG) to supply electrical power.  One RTG was available for use by New Horizons.  To accommodate this constraint, the spacecraft electronics were designed to operate on less than 200 W.  The travel time to Pluto put additional demands on system reliability.  Only after a flight time of approximately ten years would the desired data be collected and returned to Earth.  This represents the longest flight duration prior to the return of primary science data for any mission by NASA.  The spacecraft system architecture provides sufficient redundancy to meet this requirement with a probability of mission success of greater than 0.85.  The spacecraft is now on its way to Pluto with an arrival date of July 14, 2015.  Initial in-flight tests have verified that the spacecraft will meet the design requirements.

## 1    Introduction

A reconnaissance mission to Pluto and the Kuiper Belt immediately brings a number of issues to mind for those charged with the development of the spacecraft.  At a distance from the Sun of over 28 Astronomical Units (AU) at the nearest point in its orbit, the travel time to Pluto is significant and a spacecraft must be designed to be highly reliable; implying both simplicity in design and sufficient redundancy to guard against single point failures.  The solar energy at Pluto is on the order of 1/1000 of the irradiance received in Earth orbit; therefore, the spacecraft must carry its own energy source.  The only currently available technology is the Radioisotope Thermoelectric Generator (RTG) which uses the thermal energy created by the decay of Plutonium 238 to produce electrical energy.  A single RTG was available to the development team which limited the total electrical energy available to operate the spacecraft to approximately 200 Watts at the time of the Pluto encounter.  The limitation in electrical energy also required a very efficient thermal design that did not require significant electrical energy to supply make-up heat.  The communications system, as well as the command and data handling system design, is dominated by the fact that the round trip light time at Pluto is nine hours. The spacecraft is required to operate autonomously throughout the close encounter with Pluto and it will take months to return the entire data set to Earth.

With the limited power of a single RTG, both the power dissipation of each subsystem and the operation of the spacecraft were carefully considered. Design choices provided some relief such as the design of the communication system described in section 6. The remaining degrees of freedom in power management consist of the power cycling of instruments and other subsystems, and the ability to minimize the power required for thermal management. The thermal management issues are described in section 8. Though it is possible to operate with various subsystem elements and instruments on at the same time, the plan for the Pluto encounter is to operate with only those units essential to a particular observation on at any one time.

The guidance and control system design is capable of pointing the spacecraft with sufficient accuracy for the science instruments to observe Pluto, its moons Charon, Hydra and Nix, and other objects of interest during its journey. In addition, the attitude control of the spacecraft is capable of pointing the antenna toward Earth while requiring very little control activity during the journey to Pluto. For New Horizons, the guidance and control system operates in both spinning and three axis modes to meet these two requirements. Finally, the system provides the ability to make adjustments in the spacecraft trajectory to correct the residual launch trajectory errors, to correct for small disturbances encountered along the way to Pluto, and to modify the trajectory after Pluto to allow the reconnaissance of Kuiper Belt objects if NASA approves an extended mission.

The raison d'etre for the New Horizons spacecraft is to bring a selected set of science instruments sufficiently close to Pluto to gather information to meet the science requirements established by NASA in the Announcement of Opportunity AO-OSS-01 released on January 19, 2001. These requirements are described by Stern[1] and Weaver[2] in their articles in this issue. The New Horizons science team chose seven instruments for the New Horizons mission:
1. Alice[3], a UV spectrometer;
2. Ralph[4], a multi-color imager/IR imaging spectrometer;
3. REX[5], an uplink radio science instrument with radiometer capabilities;
4. LORRI[6], a long focal length panchromatic CCD camera;
5. SWAP[7], a solar wind monitor to address Pluto atmospheric escape rates;
6. PEPSSI[8], an energetic particle spectrometer to measure the composition of the ion species existing in region of Pluto; and, the
7. Venita Burney dust counter[9], which measures the density of fine dust particles in the solar system along New Horizons' trajectory from Earth to Pluto and beyond.

The placement of these instruments on the spacecraft is shown in Figure 1. The optical instruments (Alice, Ralph, and LORRI) are approximately co-aligned. The Alice instrument has a secondary optical aperture aligned with the High Gain Antenna used by REX, thereby allowing both instruments to make simultaneous measurements of Pluto's atmosphere during that portion of the mission trajectory when Pluto and Charon pass between the spacecraft and the Sun and Earth, respectively. SWAP and PEPSSI are oriented to use both the spinning and three axis control modes to collect data. The orientations are also set to minimize solar glint into the PEPSSI aperture and maximize the ability of SWAP to collect pick-up ions during the approach and close encounter at Pluto. The Venita Burney dust counter is located on the side of the spacecraft pointing in the direction of the spacecraft velocity vector (the so-called "ram" direction) for the majority of the trip to Pluto.

The development team used these design constraints while transforming the concept of Figure 1 into the spacecraft shown in Figure 2. The following sections provide additional insight about design choices, principally focusing on how the spacecraft operates.

## 2 Spacecraft Configuration

### 2.1 *Mechanical Configuration*

The mechanical configuration of the spacecraft was driven by the need to spin the spacecraft about the axis defined by the antenna assembly, maintain spin axis alignment during launch, place instruments and propulsion system thrusters to meet observation requirements and not interfere with one another, provide space for the spacecraft subsystems and accommodate the launch loads. The Ulysses spacecraft[10] had similar design requirements. The New Horizons design team used its general configuration as a starting point for the spacecraft mechanical configuration. These included:
1. A configuration that aligns the principal moment of inertia axis with the High Gain Antenna (HGA), the +Y axis of Figure 1
2. Placement of the single RTG in the X-Z plane of the spacecraft to increase the angular momentum and maximize the distance of this source of radiation from the electronics and instruments.

This configuration provides a highly stable platform to point the spinning spacecraft in a desired direction, usually toward the Earth.

The foundation of the spacecraft structure is a central cylinder located within the spacecraft main body. This single piece, machined aluminum ring forging is the focal point for the load path and as such, is the primary load bearing member of the spacecraft. The central cylinder also incorporates the separating interface adaptor to the third stage. Directly attached to the central cylinder are the propulsion system fuel tank (centered within the cylinder) and the top, aft and main bulkhead decks. The bulkhead decks provide flat mounting surfaces and interfaces for sub-systems. Five additional side and rear facing decks attach to the aft, top and main bulkhead decks. The decks are of an aluminum honeycomb type sandwich structure in which areas subject to high stress include magnesium edge members.

The titanium RTG support structure is attached to the decks on the opposite side of the cylinder from the instruments. The RTG structure is comprised primarily of titanium in order to provide high stiffness, low mass and low thermal conductivity. Since the RTG support structure was also required to provide electrical isolation, an all-metallic electrical isolation system was developed that didn't compromise structural stiffness. The isolation system employs multiple layers of non-conductive surface coatings applied to metallic plates located at the structure joints. To meet RTG interface requirements, an aluminum interface flange is attached to the titanium structure on the RTG side. Additional requirements to ensure proper stiffness and prevent deterioration of material properties due to increased RTG temperatures necessitated the addition of cooling fins to the RTG adaptor collar.

The propulsion system (hydrazine) fuel tank is centered on the axis defined by the principal axis and the center of mass of the spacecraft. At this location the variation in the fill fraction, as the fuel is used, has minimal impact on changes in the alignment of the spin axis. This ensures when the spacecraft is spinning the HGA remains pointing in a fixed direction throughout the mission's life. The tank has a capacity of 90 kg of hydrazine. The total mass at launch was 77 kg of hydrazine and helium pressurant to meet the maximum allowable launch mass and to ensure stability of the combined spacecraft and third stage stack.

The placement of the fuel tank has two further advantages. It is surrounded by the cylinder that carries the launch loads directly into the third stage attach fitting. Since the tank is between the instruments and most of the electronics and RTG, the fuel further reduces the radiation effects of the RTG. The modest radiation output of the RTG and its placement guarantees that total dose received by the electronics is less than 5K rads throughout the primary phase of the mission (Pluto data return). The majority of this dose is received at Jupiter when the spacecraft passes at a distance of 32 $R_j$ (Jupiter radii). In addition, waste heat from the RTG and the electronics are used to keep the hydrazine sufficiently above the freezing point without significant use of make-up heater power.

## *2.2 System Configuration*

Figure 3 provides an overview of the major functional system elements and their connectivity[11]. To meet the overall reliability requirements there is significant block redundancy. There are two Integrated Electronic Modules (IEMs). Each IEM contains: a Command and Data Handling (C&DH) processor; a Guidance and Control (G&C) processor; RF electronics which are the heart of the communication system; an instrument interface card which provides connectivity to each instrument; and a 64 Gbit solid state recorder. In addition there is block redundancy in much of the remaining system elements, including the TWTAs, star trackers, and Inertial Measurement Units (IMUs). Other system elements (such as the Power Distribution Unit) include redundant boards or have redundancy at the circuit level (for instance the shunt regulator unit has triply redundant control elements[12]).

To improve system reliability, significant cross-strapping below the block level is included in the design. For example, a C&DH processor in IEM 1 can be used as the controlling unit and the G&C system can operate using the G&C processor in IEM 2. The instruments' interfaces and much of their electronics (up to the focal plane detectors) are redundant and much of the redundant circuitry can be accessed from the instrument interface card in each IEM. This cross-strapping is continued with each C&DH processor being able to access each of the redundant 1553 buses that provide the data pathways to the IMUs, star trackers, and sun sensors.

Only a few elements of the system are not redundant. These include the: RTG; propulsion system tank, line filter and orifice; RF hybrid coupler; and, HGA. Each of the units has a very robust design and a long history of failure free service.

## 3 Propulsion Subsystem

The propulsion system[13] includes: twelve 0.8 N thrusters; four 4.4 N thrusters; and, the hydrazine propellant tank and associated control valves as shown in Figure 3. The titanium propellant / pressurant tank feeds the thrusters through a system filter, flow control orifice and a set of latch valves that prevent flow of the fuel until commanded to the open position after launch. Helium was selected as the tank instead of nitrogen to allow the loading of an additional kilogram of hydrazine. Blowdown curves for the system are shown in Figure 4. Measurements of tank pressure and temperatures at various points in the system allow the mission operations team to monitor system performance and the amount of fuel remaining in the tank.

The 16 Rocket Engine Assemblies (REAs) are organized into 8 sets and placed on the spacecraft as shown in Figure 5. Pairs of the 0.8 N thrusters (each thruster from a different set) are usually fired to produce torques and control rotation about one of the three spacecraft axes. One pair of the 4.4 N thrusters is aligned along the –Y spacecraft axis to provide ΔV for large propulsive events such as Trajectory Correction Maneuvers (TCMs). The second pair of 4.4 N thrusters is aligned to produce thrust along the +Y axis. These thrusters are rotated 45° in the Y-Z plane to minimize the plume impingement on the HGA dish. The net propulsive effect of these thrusters is therefore reduced. They still provide the required redundancy and the ability to generate thrust in both directions without a 180 degree rotation of the spacecraft.

The pulse duration and total on-time of each thruster is commanded very precisely, providing accurate control of the total impulse generated during a maneuver. The 0.8 N thrusters can be turned on for periods as short as 5 msec. as further described in Section 4.2. The initial propellant load was allocated between primary mission TCMs, attitude control (including science and communication operations), and primary mission margin. At the end of the primary mission, sufficient margin may allow for a possible extended mission to one or more objects in the Kuiper Belt. This allocation is given in Table 1. The original margin was augmented during the final mission preparations when the unused dry mass margin was converted to additional propellant.

Given the mass and moments of inertia at launch, the ΔV propellant cost is approximately 4.9 m/sec/kg. A change in spin rate of 5 RPM (i.e., the change from the nominal spin rate to zero RPM for 3-axis control mode) requires approximately 0.125 kg of hydrazine.

## 4 Guidance and Control

The guidance and control system uses a suite of sensors to determine the attitude of the spacecraft, the propulsion system thrusters as actuators, and one of the redundant guidance and control processors to compute the required control actions necessary to meet the commanded attitude state[14]. The attitude sensors (see Figure 3) include the Inertial Measurement Units (IMUs), the star trackers, and the sun sensors. The guidance and control system is capable of providing spin axis attitude knowledge of the spacecraft to better than +/- 471 micro-radians 3σ and spin phase angle knowledge within +/- 5.3 milli-radians 3σ. The same +/-471 micro-radians knowledge 3σ is provided for all axes when the spacecraft is in 3-axis mode.  The control algorithms must maintain the spacecraft attitude to within +/- 1024 micro-radians 3σ and the spacecraft rotational rate to within +/- 34 micro-radians/sec 3σ.

*4.1  Attitude Control Modes*

Figure 6 illustrates the various attitude control modes (3-axis, active spin, and passive spin) and the four nominal state classes (TCM, operational, earth acquisition, and sun acquisition).  TCMs can be conducted in any of the three attitude modes.  The choice of mode depends on factors such as the size of the ΔV to be imparted to the spacecraft, the desire to maintain the telemetry link while executing the burn, etc.  The various operational states depend on spacecraft activity and other constraints such as the duration of the mode, the need for conservation of fuel, the level of ground monitoring, etc.

At launch, the spacecraft was put in the Passive Spin Hibernation (PS-H) state. This state precludes any on-board attitude control and minimizes the electrical power demand.  As the name denotes, it is the primary mode used during the long cruise from Jupiter to Pluto when there will be a low level of monitoring from the ground and the spacecraft is put into "hibernation".  The Passive Spin Normal (PS-N) state also precludes any on-board attitude control activity, but does not limit other power demands.  This mode is used during significant portions of the cruise from Earth to Jupiter and at other times in the mission.  The Active Spin Normal (AS-N) state is used when an attitude maneuver is required and the spacecraft is spinning.  This state is used to maintain the spin rate at its nominal 5 rpm value, or to precess the spacecraft to a new orientation.  The 3-axis mode states allow the spacecraft to be rotated about any spacecraft axis or desired axis to point in a particular orientation.  This is used to point a particular instrument to a desired target or to scan a particular instrument field of view across a particular target (see Figure 7).

The 3 Axis Normal (3A-N) state is used during most instrument activities including instrument commissioning, engineering tests, and the Jupiter science campaign.  The 3 Axis Encounter (3A-E) state is used during the period of closest approach at Pluto where acquiring data has priority.  The difference between 3A-N and 3A-E is how the on-board autonomy system reacts to critical faults (e.g., a reboot of the C&DH processor).  In the 3A-E state, the autonomy system attempts to correct the fault while keeping the spacecraft in 3A-E state to maximize science data acquisition.  Conversely, critical faults in 3A-N state (and other states) result in the autonomy system exercising one of its two "go-safe" chains known as Earth Acquisition Go-Safe.

The Earth Acquisition Go-Safe chain transitions the spacecraft from its current mode and state into Active Spin Earth Acquisition (AS-EA) following the state transitions shown in Figure 6.  Once in Earth Acquisition mode, the autonomy system configures the spacecraft with the HGA pointed to Earth using emergency uplink (7.8125 bps) and downlink (10 bps) rates.

The second "go-safe" chain, Sun Acquisition Go-Safe, is exercised by the autonomy system as a last resort in the event that communication from the ground cannot be established with the spacecraft while in Earth Acquisition mode.  This can be the result of a corrupted on-board ephemeris (which is needed by the spacecraft to determine Earth position), failed star trackers (which are needed for inertial reference), prolonged period of command inactivity (command loss timeout), etc. In Sun Acquisition (AS-EA) mode, the autonomy system configures the spacecraft to orient the HGA towards the Sun and to send a radio signal that a critical fault has occurred on the spacecraft (a so-called "red" beacon tone). Since Sun Acquisition Mode assumes that any Earth knowledge has been lost, the Sun provides the only "guidepost" for the spacecraft pointing. Additional information on the autonomy system is provided in Section 9.

*4.2 Attitude Control Requirements and Performance*

The thrusters generate the control torques that change the attitude motion required to execute the slews to targets and to point or scan the desired instrument toward or across the target. The two major strategies of observation by the optical instruments are
1. To first point the appropriate instrument field of view (see Figure 7) and "stare" with the spacecraft rotation limited to values of less than one image pixel during the maximum expected exposure time. This rate limit was determined by the size of the LORRI image pixel and the expected maximum exposure time of 10 msec.
2. To scan about the spacecraft Z axis with a rate control sufficiently precise to allow the Ralph instrument to perform Time Delay Integration of faint images.

Both of these requirements are met if the body rates are controlled with an accuracy of 34 μrad/sec.

The thrusters are used in a pulse width modulation mode to achieve the desired control precision with the IMUs used as the fine rate control sensors. The minimum pulse width available in the system of 5 milliseconds is a value set by design based on the actuation time constants of the thruster control valves. The actual performance of the guidance and control system as measured during the commissioning tests in the spring of 2006 are given in Table 2. These numbers improve (decrease) as the feed pressure of the spacecraft propulsion system decreases.

**5  Command and Data Handling**

The Command and Data Handling (C&DH) functions: command management; science and engineering data management; timekeeping; and the capability to provide autonomous system recovery and safing, are implemented with resources within the two redundant IEMs (see Figure 3). These resources are:
- C&DH Processor Card
- Solid State Recorder (SSR) Card
- Instrument Interface Card
- Critical Command Decoder (CCD) on the Uplink Card
- Downlink Formatter on the Downlink Card

Communication between cards within the IEM is over a Peripheral Component Interconnect (PCI) backplane. Remote Input/Output (RIO) units provide temperature and voltage measurements for monitoring the health and safety of the spacecraft. These RIO units communicate with the C&DH using an Inter-Integrated Circuit ($I^2C$) bus. A MIL-STD-1553 serial bus interface is used between the IEMs to allow for redundancy and cross-strapping as well as the transfer of commands and data between G&C components and the G&C processors.

The C&DH system consists of these hardware resources and the software running on the C&DH processor. The processor in IEM 1 is designated C&DH 1 and the processor in IEM 2 is designated C&DH 2. During normal operations, the C&DH section of one IEM is fully powered and designated as prime, the so-called "bus controller" while the second unit is powered in a standby mode. During the long cruise between Jupiter and Pluto, this second unit will be powered off. The fault protection system monitors, from multiple locations, a pulse per second "heartbeat" from the bus controller. If the "heartbeat" is interrupted for a period of 180 seconds, the redundant C&DH will be powered on and take command as the bus controller. In addition, other safing actions would be taken as described in the section on fault protection and autonomy.

## 5.1 Command Management

Ground commands are received via the RF uplink card. A critical command decoder on the uplink card provides a simple, robust mechanism to receive and execute commands without the operation of the C&DH processor. These include C&DH processor resets, switching of bus controllers, and all other power switching commands. The C&DH software receives CCSDS (Consultative Committee for Space Data Systems) telecommand transfer frames from the ground via the RF system uplink card. The C&DH software extracts telecommand packets from the transfer frames. These telecommand packets are delivered to other sub-systems on the spacecraft or to C&DH itself. The telecommand packets that C&DH receives can contain commands to be acted upon in real-time or commands to be stored for later use. Commands intended for later use are designated as command macros. Command macros can consist of a single simple action (such as commanding power on or off to a subsystem) or a string of switching commands and information (data) for use by the various subsystems. Each C&DH system has 0.75 Mbytes of storage for command macros.

The C&DH system supports time-tagged rules. These trigger and execute one or more of the command macros when the value of the on-board Mission Elapsed Time (MET) is greater than or equal to the value of MET specified in the rule. Mission Operations uses time-tagged rules to implement almost all on-board activities, including science activities at Jupiter and Pluto. There is storage for 512 time-tagged rules.

The C&DH software implements the on-board autonomy system. An autonomy rule is a postfix expression whose inputs can include any spacecraft telemetry point. An autonomy rule triggers when it has evaluated to true for the required number of times in a given interval. A triggered autonomy rule also executes one or more of the command macros. There is storage for 512 autonomy rules. A further description of the autonomy system is given in Section 9.

## 5.2 Time Management

The accuracy of the correlation of Mission Elapsed Time (MET) to Universal Time (UT) is of major importance to support navigation, G&C activities, and the collection of science data. The Ultra Stable Oscillator (USO) used as the onboard source of a 1 PPS signal maintains the spacecraft time base. Careful design of the ground to MET clock register in the C&DH system ensures on-board correlation of better than +/- 4 seconds when the spacecraft is at Pluto (a 9 hour round trip light time delay). Science instrument time correlation (post-facto) requirements are +/- 10 milliseconds for REX, LORRI, Ralph and Alice. Post launch measurements during July 2006 have verified that the post-facto timing correlation is well within this requirement.

*5.3 Data Management*

The instrument interface card communicates to each of the instruments. Communication consists of:
1. Providing commands and mission elapsed time from the IEM to each instrument;
2. Collecting housekeeping data and low rate science data from the instruments to the IEM;
3. Receiving and formatting instrument high rate science data and IEM generated header data;
4. Distributing spacecraft time markers to each instrument; and,
5. Processing and transferring IEM analog telemetry voltages (data input from the RIO devices using the I2C bus).

Each instrument interface card (in IEM 1 and IEM 2) provides connections to both sides of each instrument's redundant electronics thereby providing significant cross coupling and increased reliability.

The Solid State Recorder (SSR) consists of 64 Gbits of non-volatile storage organized in 16 independently addressable segments corresponding to the 16 physical memory stacks on the SSR card. Raw science data can be streamed onto the recorder at rates up to 13 Mbits/sec. Once a segment is filled, the recorder switches to the next segment. Raw science data is read off the SSR, compressed and written back to the recorder for later transmission. Erasing of the SSR is done on a segment basis, after all data in that segment is either transmitted or compressed and stored to another segment. Once a segment is erased it is available for storage of new data. The C&DH software provides several mechanisms to control the writing and playback of data onto the recorder. The recorder uses the concept of data types to allow for differentiation of the data and thereby control it. There are a total of 51 data types, including: raw (non-packetized) high-speed science data; compressed versions of the science data, with each type of compression of a particular type of science data resulting in a different data type; data from instruments that produce low-speed, packetized science data; and, instrument and spacecraft housekeeping data.

The C&DH software also implements SSR bookmarks. Bookmarks allow for the accessing of data on the SSR associated with a specific activity. Bookmarks are opened at the start of the activity and closed at the end of the activity. Data types are included in the bookmark. Bookmarks are a substitution for using Mission Elapsed Time (MET) to define the data to be compressed or played back. All commands to compress or play back data can be specified using either bookmarks or MET. The priority of the data types to be played back from the SSR can be specified. This applies both to a playback by bookmark or a playback by MET.

The New Horizons software implements loss-less compression and lossy compression. Non-packetized science data is read off of the SSR, compressed and formed into CCSDS packets, and written back to the SSR. There is also the option to read the non-packetized science off of the SSR, form the data into CCSDS packets without doing any sort of compression, and write the data back to the SSR.

Loss-less compression can be combined with sub-framing, or windowing. Rather than performing the loss-less compression on the entire image, it is possible to specify up to eight sub-frames of the image, and then perform the loss-less compression on the data within these subframes.

A special method of sub-framing is applied to images collected by the Linear Etalon Imager Spectral Array (LEISA) IR sensor within the Ralph instrument. This is known as Dark-sky Editing. A LEISA frame consists of one spatial dimension (X) and one wavelength dimension (Y). Observations are made with the spacecraft oriented such that a rotation causes the observed object to pan across the focal plane at a rate of approximately one spatial pixel per frame. For each frame, a sub-frame can be defined such that the number of pixels which most likely contain the object are determined and stored on the SSR. A single sub-frame "walks" across a sequence of LEISA images. The original Y offset of this sub-frame, the direction it moves from image to image, and the pixel rate it moves from image to image can be specified, as well as the size of the sub-frame.

## 6   Communication System

The New Horizons RF telecommunication system[15] provides the command uplink, telemetry downlink, and essential elements of both the REX instrument and radio navigation capabilities. The system is designed to provide communications with the Earth using the Deep Space Network (DSN) in both spinning and 3-axis stabilized attitude control modes. The system (see Figure 3) is comprised of an antenna assembly, RF switch network, hybrid coupler, redundant Traveling Wave Tube Amplifiers (TWTAs), redundant Ultrastable Oscillators (USOs) and the uplink and downlink cards located in the redundant IEMs.

The antenna system includes both a forward antenna system (aligned with the +Y axis of Figure 1) and a hemispherical Low Gain Antenna (LGA) mounted below the spacecraft (not shown in Figure 1) and aligned along the –Y axis. The forward antenna assembly[16] consists of the +Y hemispherical-coverage LGA, and two co-aligned parabolic antennas. The 2.1-meter High Gain Antenna (HGA) was designed to meet a requirement for a minimum of 600 bps downlink telemetry rate at 36 AU to return the Pluto data set. The HGA provides better than 42 dBic gain for angles within 0.3 degrees of the +Y axis. The secondary reflector assembly consists of a Medium Gain Antenna (MGA) and the HGA subreflector. The MGA allows communication at larger angles between the +Y axis and Earth (up to 4 degrees), and specifically allows commands to be received by the spacecraft at ranges up to 50 AU. The two LGAs provide communication with Earth at any attitude orientation early in the mission. The LGAs are capable of maintaining communications up to distances of approximately one AU from the Earth. Beyond that distance (reached in the spring of 2006), only the MGA and HGA can be used.

## 6.1 Command Reception and Tracking

The uplink card provides the command reception capability as well as a fixed down-conversion mode for the uplink radio science experiment (REX). The uplink command receiver utilizes a low-power digital design that significantly reduces the power consumption of this critical system. Previous deep space command receivers consume ~12 watts of primary power; the digital receiver developed for New Horizons consumes approximately 4 watts. Since both receivers are typically powered on the total power savings is 16 watts, a mission-enabling achievement for New Horizons with its very limited total power budget of ~200 watts. The uplink card also provides critical command decoding, ranging tone demodulation, X-band carrier tracking, and a Regenerative Ranging subsystem. New Horizons also flies a non-coherent Doppler tracking and ranging developed by APL[17], implemented largely on the uplink card, capable of providing Doppler velocity measurements of better than 0.1 mm/s throughout the mission.

Range tracking of interplanetary spacecraft is normally accomplished by sending tones phase-modulated onto an RF carrier from the DSN to the spacecraft, receiving the signal with a wide-bandwidth spacecraft receiver and retransmitting that signal back to the DSN stations. There the retransmitted signal is processed to determine two-way round trip light time and thus the distance to the spacecraft. The primary error source in this system is the received uplink noise accompanying the ranging modulation signal on the spacecraft, which is also amplified and retransmitted back to the DSN. In addition to supporting standard tone ranging, the New Horizons communication system has incorporated a Regenerative Ranging circuit (RRC) to limit this turnaround uplink noise. This implementation uses a delay-locked loop (DLL) that generates an on-board replica of the received ranging signal and adjusts the timing of the onboard signal to align it with what the spacecraft receives. The signal transmitted to the ground is free of wideband uplink noise, significantly reducing the primary error source contribution. This will enable New Horizons to determine spacecraft range to a precision less than 10 m (1-sigma) at Earth ranges to beyond 50 AU, or achieve mission range measurement accuracy for integration times orders of magnitude shorter than those for sequential ranging.

## 6.2 Ultra Stable Oscillator Performance

The New Horizons Ultra Stable Oscillator (USO) is a critical component of the RF telecommunication system, providing the precision reference frequency (30 MHz) for the uplink, downlink, and the radio science experiment (REX). The USO is built at APL, and is based on heritage of systems developed over the last 30 years for missions such as Mars Observer, Cassini, GRACE, and Gravity Probe B. New Horizons carries two USOs. Each USO is a pristine version of an ovenized crystal oscillator. Short-term frequency stability (Allan deviation) at 1 second and 10 second intervals is better than $3 \times 10^{-13}$ and $2 \times 10^{-13}$, respectively.

This stability, and the USO low output phase noise (< -125 dBc/Hz @ 100 Hz), are crucial to the uplink radio science experiment (REX).

*6.3 Downlink System Performance*

The downlink card in each IEM is the exciter for the TWTAs and encodes frame data from the spacecraft C&DH system into rate 1/6, CCSDS Turbo-coded blocks.  It also calculates and inserts navigation counts into the frame data to support the noncoherent Doppler tracking capability.  In addition, it is used to transmit beacon tones during the hibernation cruise period.  The redundant TWTAs are the high power amplifiers for the downlink signal.  The hybrid coupler connects the downlink exciter outputs and the TWTA RF inputs.  This allows either TWTA to be connected to either downlink card.  The two TWTA outputs are connected via the RF switch assembly to the antennas.  The network allows both TWTAs to simultaneously transmit through the HGA (if there is sufficient spacecraft power), with one transmitting a Right Hand Circular (RHC) polarized signal and the other a Left Hand Circular (LHC) polarized signal.  The DSN has the capability to receive both signals and combine them on the ground to enhance the received signal-to-noise ratio and thereby increase the data rate by approximately 1.9 times above that using a single TWTA.  The resulting downlink data rate is shown in Figure 8.  The downlink system will guarantee that the entire Pluto data set (estimated to be 5 Gbits after compression) in 172 days with one 8 hour pass per day using the DSN 70 M antennas.  If there is sufficient power, such that both TWTAs can be used, the time to downlink the data set can be reduced to less than 88 days.

# 7    Power System

The power system supplies and distributes power to the spacecraft subsystems and the instruments.  It also provides hardware redundancy and fault protection.  The key components of the power system (see Figure 3) are the General Purpose Heat Source Radioisotope Thermoelectric Generator (GPHS-RTG); the Shunt Regulator Unit (SRU); the external power dissipation shunts; the Power Distribution Unit (PDU); and, the Propulsion Diode Box (PDB).  A functional summary of the power system is given in Table 3.

*7.1    Radioisotope Thermoelectric Generator Performance*

The "F8" GPHS-RTG supplied by the Department of Energy[18] is the latest in a series of RTGs of the same design supplied for NASA missions beginning in the late 1980s.  The unit converts heat generated by radioactive decay of the plutonium heat source into electricity using silicon-germanium (SiGe) thermocouples (designated "unicouples" because of the close physical proximity of the "hot" and "cold" junctions of the thermocouple).  The plutonium heat source is made up of 72 individual pellets of plutonium dioxide.  Each pellet is encased in an iridium cladding and assembled (4 to a module) into the 18 heat source modules that make up the total RTG inventory.  The RTG assembly is illustrated in Figure 9.  The fully fueled RTG had 132,465 curies of plutonium at launch with a half life of 87.7 years[19].  The thermal output of the heat modules was 3,948 watts at launch[20] which is radiated to space during flight.  Results of spacecraft RTG integration and early mission performance tests have been flawless with the "F8" normalized power output tracking above its contemporaries[21].

*7.2    Shunt Regulator Unit (SRU)*

The SRU is the power interface with the RTG.  It is designed to maintain the bus voltage to 30V by dissipating the excess RTG power in either external resistive shunts (waste heat is radiated to space) or internal spacecraft heaters depending on the thermal control needs.  The three SRU loop controllers use a majority-voted design to control the sixteen sequential analog shunts that maintain a constant load on the RTG during spacecraft operational mode changes.  Each full shunt is capable of dissipating 19.5W with the first two shunts steerable to internal heaters in half-power increments.  The shunts were designed to be n+1 redundant with a fully fueled RTG and can be individually disabled if a fault is detected.  Relays within the SRU maintain the configuration of the shunts, internal heaters, or external dissipaters. The spacecraft bus voltage is set at 30.25V which is slightly above the maximum power transfer level of the RTG to account for voltage drops as the power is distributed to the individual loads.  The SRU also contains a 33.6 mF capacitor bank that provides for short duration current surges at load turn on or fault condition.

*7.3    Power Distribution Unit (PDU) and Propulsion Diode Box (PDB)*

The PDU contains fully redundant solid-state power switching, pulsing and monitoring functions and hardware-based fault protection features for the spacecraft.  The communication interface to the IEM is through the redundant 1553 interfaces with redundant UART serial links for the passage of critical commands and telemetry.  Loads such as the IEM, command receiver, USO, and the PDU 1553 board are configured as critical loads with both the primary and redundant units powered on by default.  Software and hardware enables are required to turn off any of the spare units.  Dual-level low-voltage sensing sheds non-critical loads if the bus voltage drops below the set points (nominally 28.5 volts).  Additional bus protection features within the PDU also address the power limited nature of the RTG power source.  The circuit breakers have selectable levels for tight load monitoring and allow multiple attempts at power cycling the loads.  Loads are fused as a final protection measure.  The PDU also contains spacecraft configuration relays, sun sensor interface, sensing of load telemetry (both voltage and current) and temperature telemetry from the power system.  The PDB is the power system interface between the PDU and the propulsion thrusters, latch valves, and the catalyst bed heaters.

Figure 10 provides the power for each of the "nominal" operating modes.  The spacecraft power requirements vary as different subsystems are powered to perform various functions.  The requirement is smallest during the cruise mode when most instruments and some subsystems such as the transmitter and IMUs are powered off.  Data return after the Pluto encounter is performed in a spinning mode.  The downlink power assumes only one transmitter on.  If the second TWTA is powered to increase the data rate downlink, as discussed in Section 6, the spacecraft must be in a passive spin state so that other components, such as the catalyst bed heaters, can be powered off to accommodate the additional 31W.  TCM operations can be performed in several ways (see Section 4.1); the minimal power mode shown here is the active spin mode. Science operations at Pluto are performed in the 3 axis control mode with only those instruments powered that support a particular observation (the maximum shown here). The two safe modes (Earth Safe and Sun Safe) require the propulsion system's catbed heaters to be powered, thus requiring the same power as a TCM. Figure 11 illustrates the expected power provided by the RTG over the lifetime of the mission and the required power for each of these modes relative to that power. Thus, New Horizons can expect to operate with significant power margin for the Pluto encounter and adequate margin for a KBO encounter at up to 50 AU.

## 8 Thermal Management

The New Horizons thermal design balances the power and waste heat provided by the RTG and the heat loss through the thermal blankets, the instrument apertures and control mechanisms to ensure each of the system elements remain within safe operating temperatures through the mission. The control mechanisms utilize thermal louvers on the lower deck of the spacecraft and the ability to dissipate excess electrical power either internally or externally. The avionics are contained within the "thermos bottle" like core of the spacecraft. The average internal temperature varies from slightly under $50^{o}\,C$ (during early operations with the lower deck facing the Sun at 1 AU) to sufficiently above $0^{o}\,C$ to ensure the hydrazine propellant does not freeze. The propulsion system components are thermally tied to the spacecraft bus and are kept warm through thermal contact with the structure. Heat leaks through the rocket engine assemblies (REAs) are sufficient to ensure they remain at quiescent operating temperatures.

The design uses approximately 15W of waste heat from the RTG to support the internal temperature. The blankets are of a sufficiently high thermal resistance to maintain internal temperatures above $5^{o}C$ using only 100W of internally dissipated electrical energy. The thermal louvers actuate if the internal temperature exceeds $25^{o}\,C$ and keeps the internal temperature from getting too high during period where the internal dissipation reaches its maximum design level.

## 9 Autonomy & Fault Protection

The New Horizons mission is long. The primary science goal can only be achieved after a nine and one half year journey culminating in a complex set of observations requiring significant time to transmit the data to Earth. Much thought and energy was devoted to fault protection during spacecraft development. This effort continues as the operations team evaluates in-flight mission performance. The fault protection architecture uses the redundancy of the spacecraft system as shown in Figure 3 if off-nominal operation is detected. Basic elements of fault protection are resident in redundant elements of the PDU. The PDU monitors C&DH bus traffic and will automatically switch to the alternate C&DH system if it detects nominal C&DH processor activity of the controlling system has stopped. The major elements of fault protection are implemented by software running on the controlling C&DH processor. This software is the principal component of the autonomy subsystem. The software evaluates telemetry data in real time and, based on the evaluation will take one or more of the following actions:
1. execute a set of commands to correct a detected fault;

2. generate a "beacon tone" to alert operators that an event on the spacecraft requiring attention has occurred; or,
3. execute one of two "go safe chain" command sets which puts the spacecraft into either an Earth Safe or Sun Safe state as described in Section 4 (in the event of a critical fault).

The evaluation of on-board data is performed by a set of "rules" that check for data that exceed defined limits for a period of time. The time period (or persistence) of the exceedance varies from rule to rule. The persistence length minimizes the chance of a rule "firing" on noisy data, or on transient data that occurs during a commanded change in spacecraft pointing. Processors (other than the C&DH processor whose activity is monitored by the PDU circuitry) are monitored via a set of "heartbeat" rules that use a telemetry point to determine if the processor is stuck either at a "one" or "zero" state. The persistence of each of these "heartbeat" rules is adjusted as appropriate to match the nominal operation of the specific processor. The autonomy software can also compute dynamic limits. For example, the autonomy system monitors the propulsion system for potential propellant leaks. The system monitors the propellant as a function of both the fuel tank pressure and temperature using the ideal gas law to compute a current volume and compares it to an initial value set at a previous time appropriate to the phase or mode of spacecraft operation. At the time of launch, the autonomy system used 126 rules to determine the state of health of the spacecraft.

The command sets are organized as user defined macros and stored in memory space defined by the C&DH system. The macros can include any allowable C&DH command and can be used to power units on or off, change spacecraft modes, enable or disable autonomy rules or execute other macros. These macros can be executed by either real time commands or by the autonomy subsystem. The macros can also be executed by time-tagged commands, allowing the commands in the macros to execute at a specific time in the future. The autonomy subsystem used 132 macros at launch. This set has been modified as the spacecraft position along its trajectory has changed and will continue to be modified as different phases of the mission occur, system performance changes, and operational experience dictates.

The capabilities of the autonomy system are used to support a number of mission operation tasks as well as providing fault protection. For example, the "command load" sequences generated by the mission operations team are loaded into one of two memory segments. Upon the completion of one sequence, an autonomy rule is used to switch to the other sequence. The autonomy rules also check to see that an appropriate sequence has been loaded into the second memory segment and if it has not, a rule fires causing the system to enter the "go safe" chain and point to Earth.

## 10  Performance and Lifetime

For mission success, the New Horizons spacecraft must continue to function through post-Pluto data playback. As discussed above, the system design included numerous element redundancies to increase the probability of proper operation beyond that of a single string design. A Probabilistic Risk Assessment (PRA) was carried out to guide overall system architecture, with initial results given at CDR and final results several months prior to launch. The results of the final PRA are shown in Figure 12 for the best and worst case launch dates and mission lifetimes for minimum science and full science. The bands in Figure 12 give the 90% confidence limits for each calculation, with the mean shown as the vertical line in each distribution. The mean probability of success remains above the design goal of 85% for the actual mission duration, and is acceptable for all possible mission durations based on changes in launch date.

Prior to launch, the sensitivity of the mean probability of success to changes in failure rates for individual system elements was examined, in part to understand the effect of unresolved anomalies from integration, test and prelaunch activities on the probability of success. In addition, this sensitivity analysis studied how the probability of success estimates depend on the accuracy of the failure rates for individual system elements. The results from altering the failure rates fall well within the uncertainty range for the PRA results in Figure 12, indicating that the impact of unresolved issues or inaccurate element failure rate estimates on the overall mission reliability was negligible.

Management of the "limited life" system elements as well as use of the consumable resources on-board the spacecraft are essential to achieving these lifetime probabilities. Limited life items are given in Table 4. The limited life items are discussed in three categories:
1. Items exercised very few times under normal operations (e.g., Latch Valves, RF Switches);
2. Items that are purely time dependent (e.g., IMU leak rate); and,
3. Items used regularly under normal operations that must be carefully managed (thruster valve cycles, TWTA and IMU on-time, and, IEM and SSR flash cycles).

The latch valves that supply attitude control thrusters were commanded to the open state shortly after launch and are not planned to be operated again. The latch valves that supply delta-v thrusters are only opened for trajectory correction maneuvers, resulting in a very small number of cycles. RF switches are operated to check out the redundant elements of the RF system. Some operations remain as the communication link uses the MGA or HGA as the mission continues, but these operations are few in comparison to the lifetime capabilities of the switches.

The IMU leak rate was measured before launch and was determined to be substantially small such that it provided margin greater than ten times the requirement. A gross change in leak rate is measured by means of the telemetry system. No change in the rate has been observed during the first year of flight.

The TWTA filaments are rated at 260,000 hours. The planned operating time through the Pluto downlink is less than 45,000 hours. Thus there is sufficient margin in the use of a single TWTA to meet mission requirements.

The IMU operating hours are limited by the gyro laser(s) operating life. These units were carefully selected during manufacturing to ensure lifetimes for all three axis measurements exceeding 42,000 hours. Pre-launch estimates of the projected lifetime actually exceeded 52,000 hours. These units operate a significant amount of time during the early cruise to Jupiter and during the Jupiter encounter. After Jupiter, the spacecraft spends most of the next seven years in the hibernation mode during which the IMUs are off. By this means, the operating time on each IMU is planed to be less than 23,000 hours.

The thruster valve cycle limit of 273,000 was adopted from the limit used on the Cassini program. This limit was derived from qualification tests, which demonstrated performance beyond 409,000 cycles. New Horizons uses the same thruster as the Cassini and Voyager missions. The Voyager spacecraft, which experienced its only failure at over 500,000 cycles, provides valuable in-flight data to increase confidence in the thruster valve cycle performance. New Horizons plans to cycle each thruster valve approximately 200,000 times, well under the 273,000 cycle limit.

The New Horizons spacecraft is fully functional with significant margins for both consumable and life limited resources to support the full mission requirements at Pluto. The initial cruise phase and Jupiter encounter provides many opportunities to fully exercise the system elements and to allow the mission operations team to demonstrate the kind of operational activities planed for the Pluto reconnaissance. With the excellent launch and the power supplied by the RTG, the spacecraft is expected to operate productively well into the Kuiper Belt providing NASA and the Science Community the opportunity to explore this region after the 2015 rendezvous is completed.

Acknowledgements

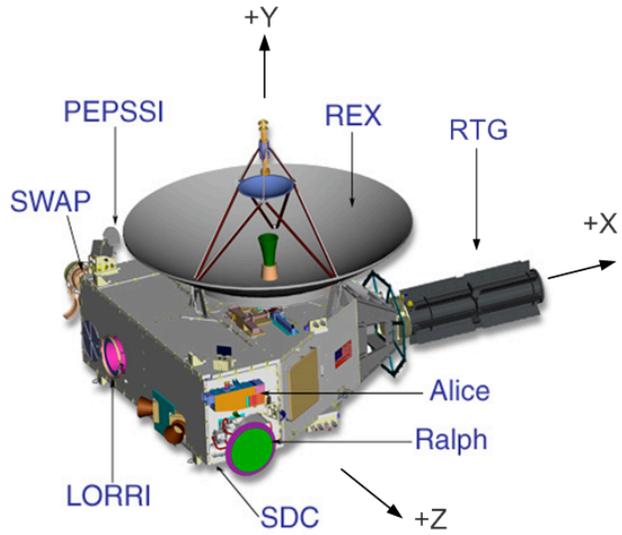

Figure 1: The New Horizons spacecraft supports seven science instruments and is powered by a single RTG.

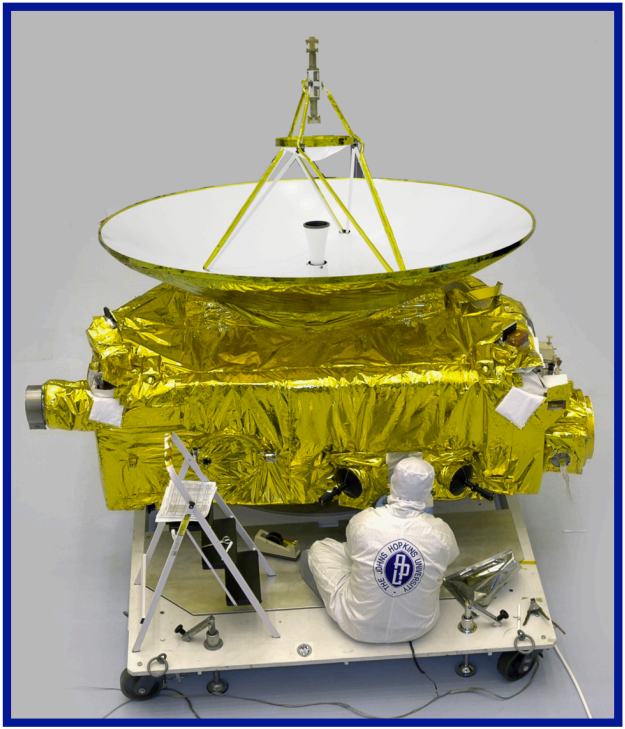

Figure 2: The New Horizons spacecraft operates in either spin mode or under three axis control. For the majority of its journey to Pluto, it spins about the +Y axis to which the antenna assembly (+Y Low Gain, Medium Gain, and High Gain) are aligned.

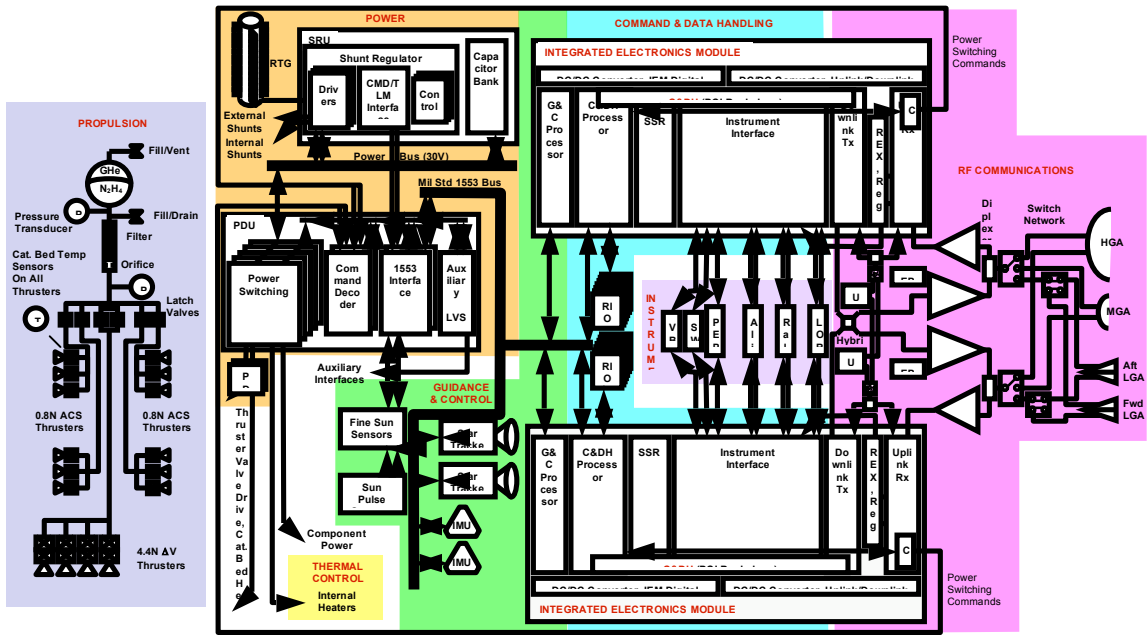

Figure 3: The New Horizons spacecraft is designed with significant redundancy to ensure the mission requirements are met after a nine and one-half year flight to Pluto.

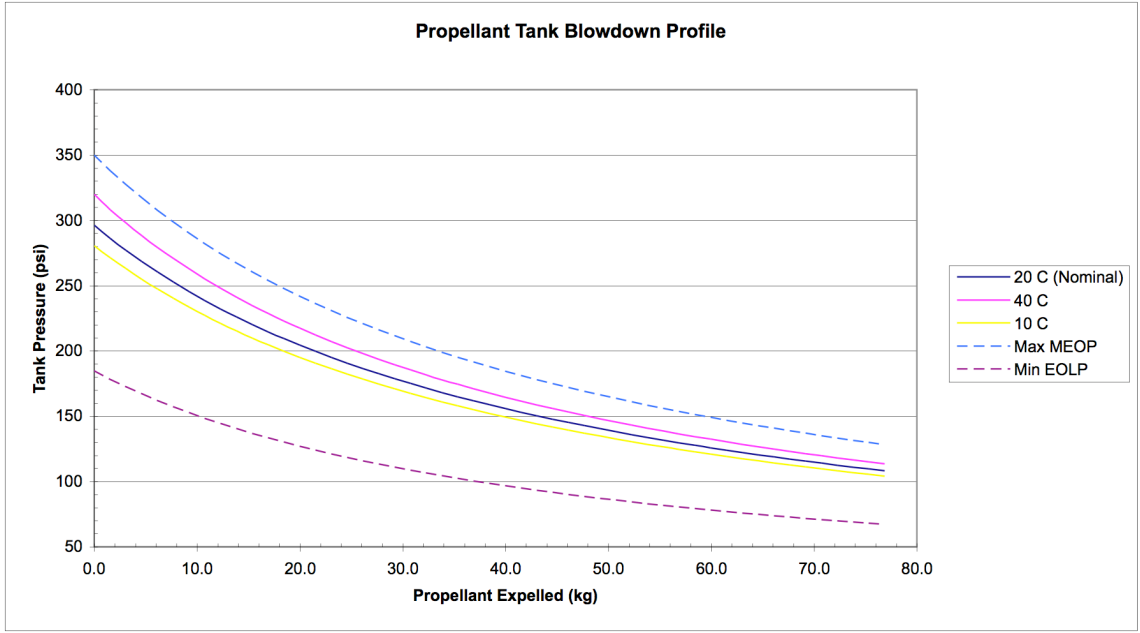

Figure 4 New Horizons propulsion system blowdown profile

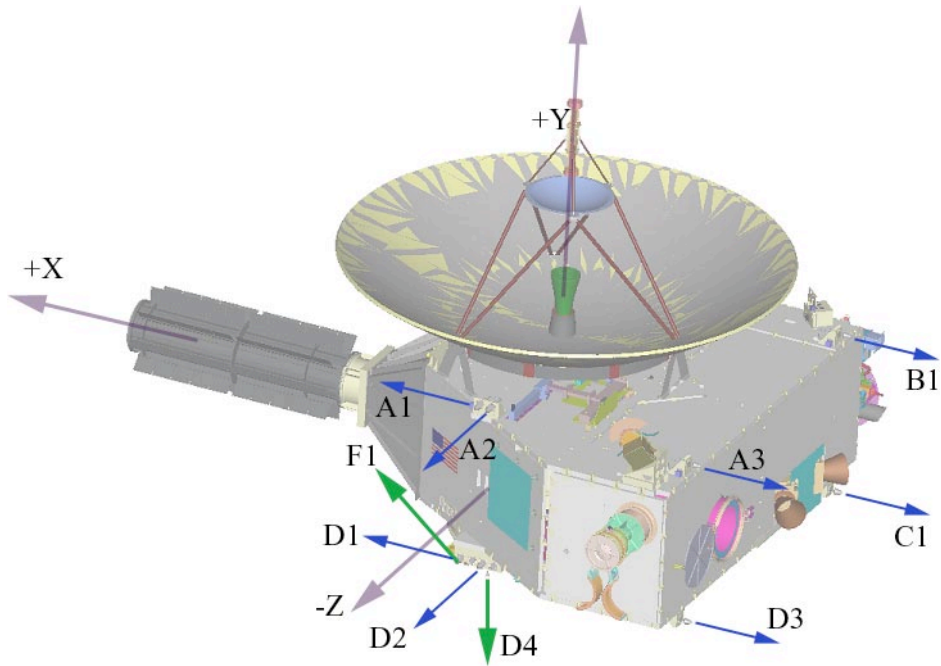

Figure 5 The propulsion system thrusters are organized in eight sets (6 of which are shown in the figure). The sets are distributed such that firing of pairs of 0.8 N thrusters located on opposite sides of the spacecraft produce the required couples for attitude control. The 4.4 N thrusters are paired in two sets on the –Y/+Z spacecraft edge (F1 & D4 shown) and the –Y/-Z spacecraft edge (not shown) and are used when larger ΔV maneuvers are required.

|  | ΔV | Propellant Usage |
|---|---|---|
| Primary Mission TCM | 110 m/s | 22.3 kg |
| Attitude Control |  | 29.3 kg |
| Primary Mission Margin | 132 m/s | 25.2 kg |
| *Original Margin Allocation* | *91 m/s* | *17.5 kg* |
| *Additional Margin Obtained from Unused S/C Dry Mass Allocation* | *41 m/s* | *7.7 kg* |
| **Total Navigation DV** | **242 m/s** |  |
| **Total Propellant Load** |  | **76.8 kg** |

Table 1 New Horizons propellant budget allocations

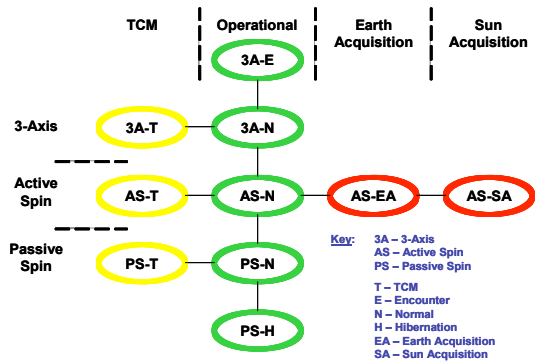

Figure 6 The guidance and control system provides three operating modes to support TCMs, nominal operations, and two safe states (Earth and Sun acquisition).

## Staring Images ~ 10 msec
## (LORRI, Ralph, Alice)

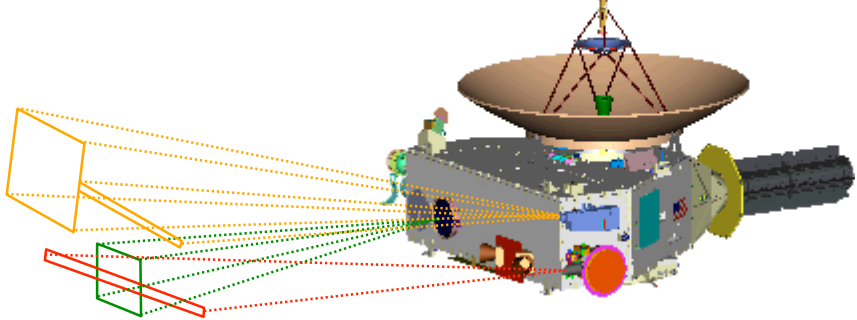

## Scanning Images~ 1 min
## (Ralph)

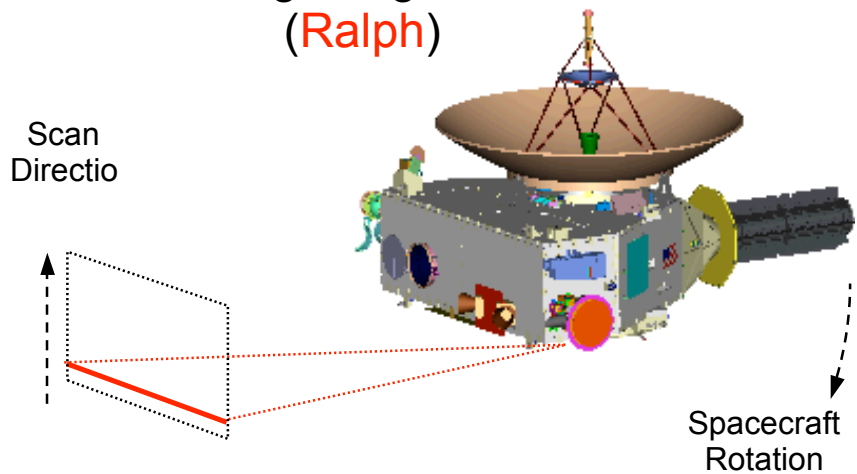

Figure 7: The optical instrument (Ralph, Alice, and LORRI) are aligned such that their fields of view nearly overlap. (Figure 7A) Only small motions of the spacecraft are needed to image the same area. Precise control of the spacecraft attitude is needed both for staring and to provide smooth scanning motion about the +Z axis for Ralph (Figure 7b).

| Thruster Combination | Average Rate Change from 5 msec pulse |
|---|---|
| +X Axis (A2) | 23 μrad/sec |
| -X Axis (B2) | 16 μrad/sec |
| +Y Axis (C1/D1) | 25 μrad/sec |
| Y Axis (C3/D3) | 23 μrad/sec |
| +Z Axis (A1/D3) | 15 μrad/sec |
| -Z Axis (A3/D1) | 16 μrad/sec |

Table 2: The spacecraft G&C system controls the body rates using small torques provided by thruster pairs on the spacecraft. The resulting minim control body rates, as measured after launch, are provided by the thruster pulse duration of 5 msec.

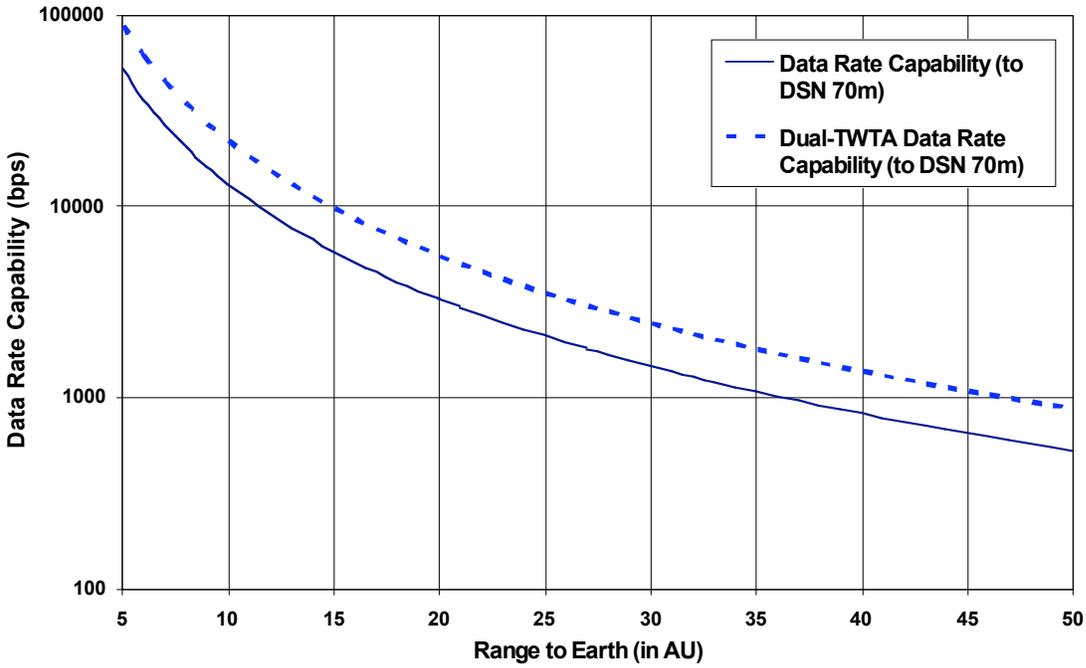

Figure 8: The RF system provides down-link telemetry above 1kb/sec at Pluto and above 700 bits/sec at 50 AU. A mode using both the redundant TWTAs at one time and combining the signals on the ground increases the data rate such that over 900 bits/sec can be transmitted at 50 AU.

| | |
|---|---|
| RTG Beginning of Mission Power | 245.7 We |
| Power Shunt Rating | 312 W |
| Regulated Bus Voltage | +30V +/- 1V |
| Switched Loads | 86 (12) |
| Thruster/Latch Valve Drivers | 56 (4) |
| Pulsed Loads | 148 (16) |
| Opto-Isolated Contact Closures | 8 (8) |
| S/C Configuration Relays | 24 (1) |
| Digital Telemetry (kBytes) | 1.22 |
| Analog Telemetry (kBytes) | 0.58 |

Table 3. Spacecraft Power System Summary. The numbers denoted as (#) are the spare services within the power system at launch.

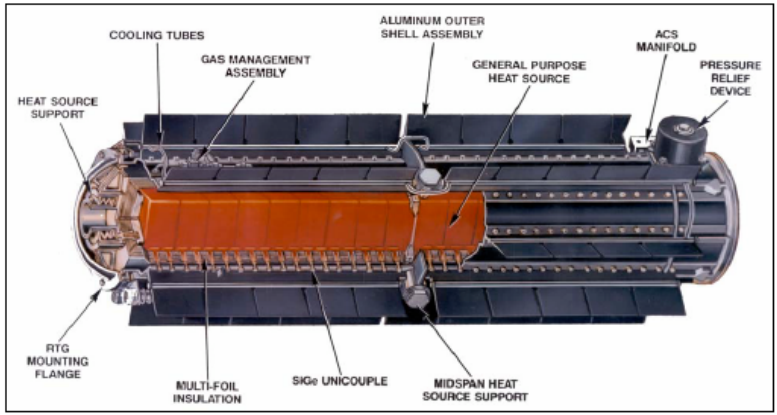

Figure 2.1-1. Components of the GPHS-RTG

Figure 9  The GPHS-RTG has 9.75 Kg of plutonium dioxide distributed in 18 General Purpose Heat Source Modules.  The modules are designed to contain the plutonium in case of a launch accident.  The aluminum outer assembly radiates the heat from the modules into space.  The unicouples generate electricity based on temperature differential between the modules and the outer assembly.  Prior to launch the assembly is filled with an inert gas to keep the unicouples from oxidizing.  Once in space, the gas pressure relief device allows the gas to escape and the RTG then operates at maximum efficiency.  (Image courtesy of the Department of Energy)

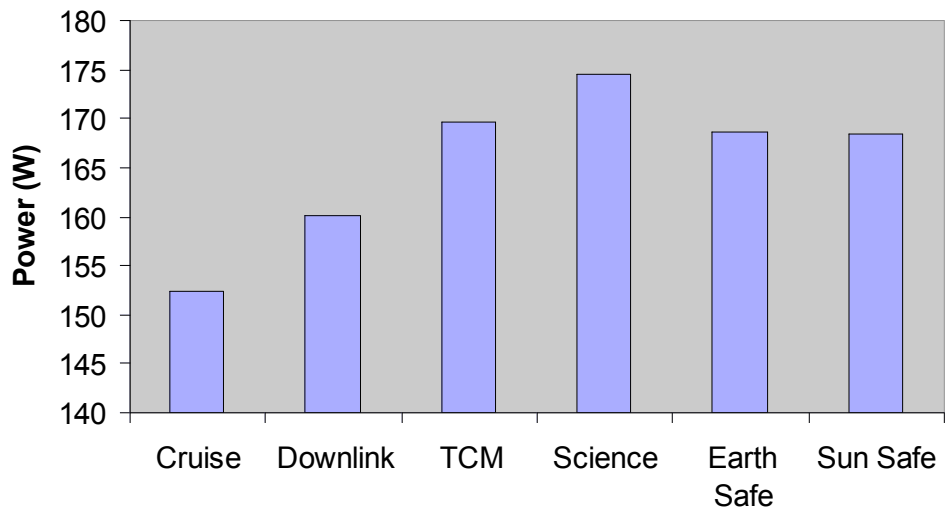

Figure 10 Spacecraft and instrument power for each of the operating states were tailored to ensure safe power margins through the primary mission.

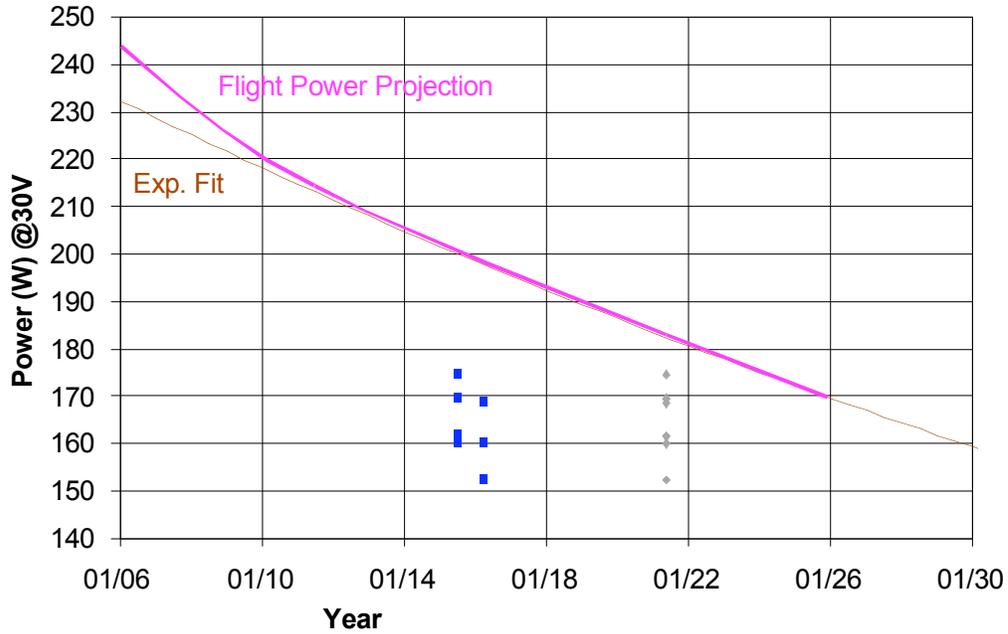

Figure 11: The Radioisotope Thermoelectric Generator (RTG) power output decreases with time primarily due to the Plutonium decay half-life of 87 years. The generator supplies sufficient power even at 50 AU (~ 2021) to support all necessary spacecraft (squares for Pluto and diamonds for a 50 AU Kuiper Belt object) modes for an encounter and data playback.

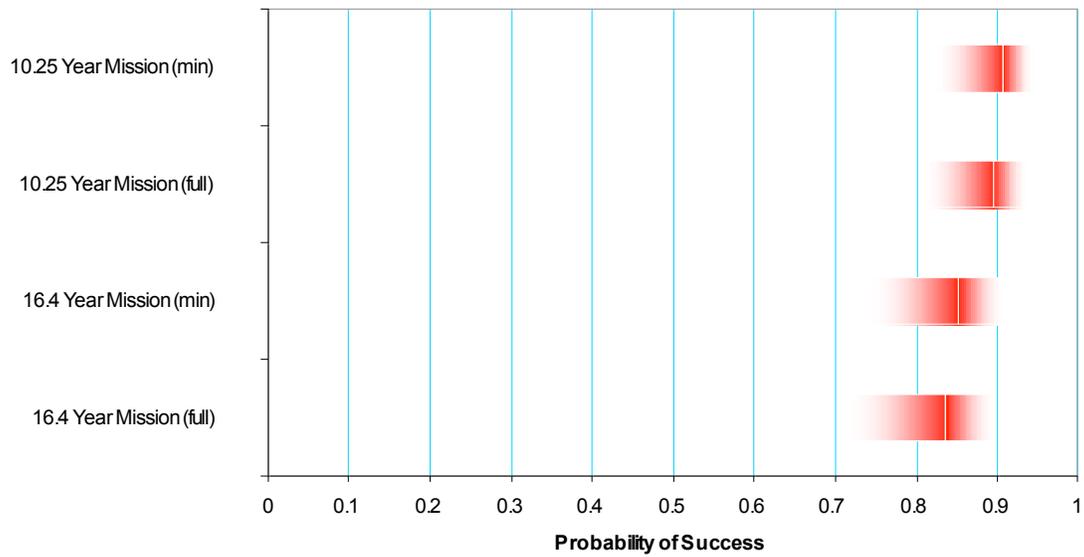

Figure 12: An analysis was performed to determine the probability of the New Horizons system design to successfully perform the mission and meet the science objectives. With the successful launch early in the 2006 launch window, the probability of meeting the full mission requirements are well above the mission requirements and the probability of mission success in an extended mission (through the Kuiper Belt to at least 50 AU) is also very high.

| |
|---|
| Latch Valve Cycles |
| Thruster Valve Cycles |
| Telecom System RF Switch Cycles |
| TWTA filament on-time |
| IMU leak rate |
| IMU on-time |
| IEM Processor Flash Cycles |
| SSR Flash Cycles |

Table 4  New Horizons limited life items

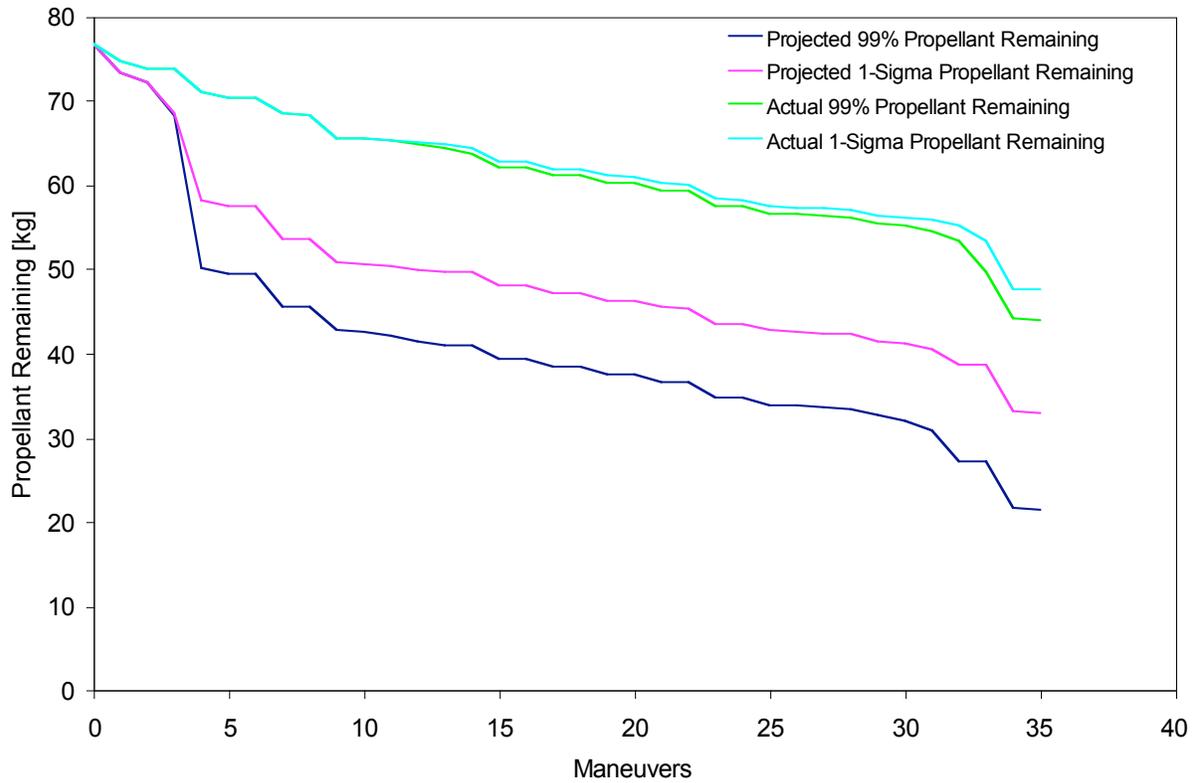

Figure 13: The hydrazine propellant is the only on-board consumable for the mission. Consumption analysis included 35 events through Pluto data return that required the expenditure of hydrazine. The pre-launch prediction indicated that a good margin would remain available (> 20 Kg with a 99% probability). Because of the excellent trajectory provided by the launch vehicle that number has increased to over 40 Kg. (Figure courtesy of Stewart Bushman)